\documentclass[12pt]{article}
\usepackage{graphicx}
\begin{document}
\baselineskip=5.5mm
\newcommand{\be} {\begin{equation}}
\newcommand{\ee} {\end{equation}}
\newcommand{\Be} {\begin{eqnarray}}
\newcommand{\Ee} {\end{eqnarray}}
\renewcommand{\thefootnote}{\fnsymbol{footnote}}
\def\a{\alpha}
\def\b{\beta}
\def\g{\gamma}
\def\G{\Gamma}
\def\d{\delta}
\def\D{\Delta}
\def\e{\epsilon}
\def\k{\kappa}
\def\l{\lambda}
\def\L{\Lambda}
\def\t{\tau}
\def\om{\omega}
\def\Om{\Omega}
\def\s{\sigma}
\def\lg{\langle}
\def\rg{\rangle}
\noindent
\begin{center}
{\Large
{\bf
Out-of-equilibrium dynamics in a gaussian trap model}
}\\
\vspace{0.5cm}
\noindent
{\bf Gregor Diezemann} \\
{\it
Institut f\"ur Physikalische Chemie, Universit\"at Mainz,
Welderweg 11, 55099 Mainz, FRG
\\}
\end{center}
\vspace{1cm}
\noindent
{\it
The violations of the fluctuation-dissipation theorem are analyzed for a trap
model with a gausssian density of states. 
In this model, the system reaches thermal equilibrium for long times after a
quench to any finite temperature and therefore all aging effect are of a transient
nature.
For not too long times after the quench it is found that the so-called
fluctuation-dissipation ratio tends to a non-trivial limit, thus inicating the
possibility for the definition of a time scale dependent effective temperature. 
However, different definitions of the effective temperature yield distinct
results. 
In particular plots of the integrated response versus the correlation function
strongly depend on the way they are constructed. 
Also the definition of effective temperatures in the frequency domain is not
unique for the model considered.
This may have some implications for the interpretation of results from computer
simulations and experimental determinations of effective temperatures.
}

\vspace{0.5cm}
\vspace{1cm}
\section*{I. Introduction}
The out-of-equilibrium dynamics of glasses and disordered systems has been
studied quite intensively in the last decade, see e.g.\cite{PCKM98}.
One reason for the interest in this area lies in the fact that glassy systems
usually do not reach thermal equilibrium within experimentally accessible times
when cooled to low temperatures. 
In such a situation the question as to whether there is a possibility for
characterizing the non-equilibrium state of the system naturally arises. 
In particular, it would be extremely helpful if some of the tools of statistical
physics could be extended to non-equilibrium situations. 
One of the efforts of quantifying the deviations from thermal equilibrium is
concerned with extensions of the fluctuation-dissipation theorem (FDT) relating
the response of the system to the two-time correlation function. 
In equilibrium, this relation is determined by the thermodynamic temperature. 
For non-equilibrium systems, the deviations from FDT can in some cases be used
for the introduction of a time scale dependent effective temperature
$T_{\rm eff}$\cite{CKP97}. 
A variety of model-calculations and numerical simulations have been performed to
analyze the behavior of $T_{\rm eff}$ obtained this way, for reviews
see\cite{CR03,barrat03}.
Recently, also some experimental investigations of the violations of the FDT have
been performed, see e.g.\cite{FDT.Exp}.
One typically finds a behavior in which $T_{\rm eff}$ exceeds the bath temperature
by a certain amount.  
This makes sense as the typical preparation of the system is given by a quench
from a high temperature into a glassy phase. 

There are, however, still some open questions regarding the usefulness of the
interpretation of the effective temperatures obtained from FDT-violations. 
For instance, for a meaningful definition of a temperature, $T_{\rm eff}$ should
be independent of the dynamical variable used for its calculation.
In case that so-called neutral variables\cite{FS02} are used, $T_{\rm eff}$ 
often is found to be independent of the variable in some long-time limit. 
Another question is concerned with the relaxation of $T_{\rm eff}$ to the bath
temperature when the system reaches thermal equilibrium on some time scale. 
This point has been investigated in some recent papers\cite{G56,G59,G60} with the
result that some care has to be taken in the determination of $T_{\rm eff}$. 

In the present paper, I will consider another model in which the system reaches
equilibrium for long times after a quench, namely a trap model with a Gaussian
distribution of trap energies\cite{MB96}. 
The same model has been introduced earlier as an approximation to a random walk
model\cite{Jeppe95} and was termed 'energy master equation' then. 
It was shown in ref.\cite{Jeppe95} that this model is able to reproduce some
features that are observed in the relaxation of supercooled liquids and glasses. 
Furthermore, recently it was demonstrated that some results of simulations on
models of supercooled liquids can (at least partly) be interpreted in terms of a
Gaussian trap model\cite{DRB03, DH03, HDS05}. 
Some features of the aging behavior of the populations of the traps have already
been discussed in ref.\cite{G56}.
There it was shown that after a quench from an (infinitely) high temperature the
distribution of populations as a function of the time elapsed after the quench
first narrows and then broadens again, in qualitative accord to some observations
in computer simulations\cite{SVS04}.

In the following Section, I will briefly recall the derivation of the
fluctuation-dissipation relations for trap models\cite{Ritort03,Sollich03,G54}. 
In Section III the results of model calculations are presented and discussed. 
The paper closes with some conclusions in Section IV.
\section*{II. Fluctuation-dissipation relations for trap models}
The derivation of the fluctuation-dissipation relations for arbitrary
continuous-time Markov processes has been given in ref.\cite{G54}. 
Usually one assumes a situation in which the system under consideration is
quenched from a high temperature $T_0$ to a low temperature $T$ in the beginning
of the experimental protocol. After a time $t_w$, called the waiting time, has
elapsed after quench, some dynamical quantities are monitored. 
In the present context, the correlation function of some dynamical variable
$M(t)$, 
\be\label{C.t.tw}
C(t,t_w)=\lg M(t)M(t_w)\rg=\int\!\!d\e\!\int\!\!d\e'
M(\e)M(\e')G(\e,t|\e',t_w)p(\e',t_w)
\ee
and the response to a field applied at time $t_w$, $H(t)=H\d(t-t_w)$, conjugate to
$M$, 
\be\label{R.def}
R(t,t_w)=\left.{\d\lg M(t)\rg\over\d H(t_w)}\right|_{H=0}
\ee
are of particular importance.
In the above expressions, the values of $\e$ are meant to represent the trap
energies. 
$M(\e)$ is the value the dynamical variable $M$ acquires in the trap with energy
$\e$ and $G(\e,t|\e',t_w)$ is the conditional probability to find the system in
trap '$\e$' at time $t$, provided it was in trap '$\e'$' at time $t_w$. 
The populations $p(\e',t_w)$ evolve from the initial populations 
$p(\e,t\!=\!0)$, typically equilibrium populations at the starting
temperature $T_0$, via $p(\e,t_w)=\int\!d\e'G(\e,t_w|\e',0)p(\e',0)$. 
For a stationary Markov process, the conditional probability $G(\e,t|\e',t_w)$ is
time-translational invariant,
$G(\e,t|\e',t_w)=G(\e,t-t_w|\e',0)\equiv G(\e,t-t_w|\e')$, and obeys a master
equation\cite{vkamp81}:
\be\label{ME}
{\partial\over\partial t}G(\e,t|\e')=
-\int\!\!d\e''W(\e''|\e)G(\e,t|\e')+\int\!\!d\e''W(\e|\e'')G(\e'',t|\e')
\ee
with $W(\e|\e')$ denoting the rates for a transition from $\e'$ to $\e$. 
In order to calculate the linear response to a field one has to fix the dependence
of the transition rates on the field $H$. 
Following Ritort\cite{Ritort03}, I use the following form of multiplicatively
perturbed transition rates: 
\be\label{kap.h}
W^{(H)}(\e|\e')=W(\e|\e')e^{\b H \left[\g M(\e)-\mu M(\e')\right]}
\ee
where $\g$ and $\mu$ are arbitrary parameters and $\b=1/T$ (with the Boltzmann
constant set to unity).
If $\mu+\g\!=\!1$ holds additionally, then the rates $W^{(H)}(\e|\e')$ obey
detailed balance also in the presence of the field, provided the $W(\e|\e')$ do
so. 
In all following calculations the unperturbed transition rates will be chosen
according to:
\be\label{W.eps}
W(\e|\e')=\eta(\e)\k(\e')\quad\mbox{with}\quad\k(\e')=\k_\infty e^{\b\e'}
\ee
as is usual for the trap model\cite{MB96,Jeppe95,trap}. Note, however, that also
other choices have been considered, see e.g.\cite{BM95}. 
As already mentioned above, I will only consider the trap model with a Gaussian
density of states, $\eta(\e)\!=\!{1\over\sqrt{2\pi}\s}e^{-\e^2/(2\s^2)}$. 
In this case, the system reaches equilibrium for long times regardless of the
initial conditions.
The equilibrium populations are given by 
$p^{\rm eq}(\e)=\lim_{t\to\infty}G(\e,t|\e_0,t_0)
={1\over\sqrt{2\pi}\s}e^{-(\e-{\bar\e})^2/(2\s^2)}$ with the temperature
dependent mean energy ${\bar\e}=-\b\s^2$\cite{Jeppe95}. 
Note that the variance $\s$ is temperature independent.

The details of the calculation of the linear response according to
eq.(\ref{R.def}) have been presented in refs.\cite{Ritort03,Sollich03,G54}. 
As a result, it is found that for mean-field trap models with a dynamical variable
$M$ that obeys a distribution with zero mean, $\lg M\rg\!=\!0$, and unit variance,
$\lg M^2\rg\!=\!1$, the fluctuation-dissipation relation can be written as:
\be\label{FDT.gen}
R(t,t_w)=\b\left[\g{\partial \Pi(t,t_w)\over\partial t_w}
		 -\mu{\partial \Pi(t,t_w)\over\partial t}\right]
\ee
Here, the correlation function 
\be\label{Pi.t.tw}
\Pi(t,t_w)=\int\!\!d\e e^{-\k(\e)(t-t_w)}p(\e,t_w)
\ee
gives the probability that the system has not left the trap occupied at
$t_w$ in the following time interval $(t-t_w)$\cite{MB96}.

In equilibrium, all quantities are time-translational invariant and one has
\be\label{fdt.eq}
R_{\rm eq}(t)=-\b(\g+\mu){d \Pi_{\rm eq}(t)\over d t}
\ee
which for $\mu\!=\!1-\g$ is just the well known FDT. 
Note that for the present choice of the transition rates, eq.(\ref{W.eps}), the
system always reaches equilibrium for $t_w\to\infty$. 
Therefore, all aging effects are of a transient nature. 
This behavior is similar to what one expects for structural glasses cooled not
too low below the calorimetric glass transition temperature.
In the present paper, the quenches are performed in the following way:
In the beginning, the system is prepared in thermal equilibrium at the
initial temperature $T_0$, i.e. 
$p^{\rm eq}_{T_0}(\e)
={1\over\sqrt{2\pi}\s}e^{-(\e-{\bar\e_0})^2/(2\s^2)}$ with
${\bar\e_0}=-\b_0\s^2$. 
The following time evolution is calculated at the working temperature $T$, 
$p_T(\e,t_w)=\int\!d\e'G_T(\e,t_w|\e',0)p^{\rm eq}_{T_0}(\e)$ where
$G_T(\e,t_w|\e',0)$ is the solution of eq.(\ref{ME}) with the transition rates
(\ref{W.eps}) evaluated for $\b=1/T$.

The violations of the FDT in out-of-equilibrium situations can be characterized
by the so-called fluctuation-dissipation ratio (FDR): 
\be\label{X.def}
X(t,t_w)={T R(t,t_w)\over \partial_{t_w}\Pi(t,t_w)}
\ee
which also allows the definition of a time scale dependent effective temperature 
$T_{\rm eff}(t,t_w)=T/X(t,t_w)$, the detailed behavior of which has been discussed
for a variety of models\cite{CR03}. 

Another way to define an effective temperature has been proposed and applied in
experimental determinations of FDT-violations\cite{CKP97,FDT.Exp}:
\be\label{Teff.om}
T_{\rm eff}(\om,t_w)={\om C(\om,t_w)\over \chi''(\om,t_w)}
\ee
relating the Fouriertransform of the correlation function (spectral density) to
the dissipative part of the susceptibility.
Of course, one would assume the various definitions to yield identical results. 
In the next Section, different definitions will be compared and the differences
will be discussed for the Gaussian trap model.
\section*{III. Results and Discussion}
I will start with a brief discussion of the equilibrium properties of the
correlation function function $\Pi(t,t_w)$. 
As already mentioned above, the system reaches equilibrium for long times after a
quench has been performed, i.e. in the limit $t_w\to\infty$.
According to eq.(\ref{Pi.t.tw}), in this limit one has for $\t=(t-t_w)<\infty$:
\be\label{Pi.eq}
\Pi_{\rm eq}(\t)=\int\!\!d\e e^{-\k(\e)\t}p^{\rm eq}(\e)
\ee
This function is plotted in Fig.1 as a function of $\t/\t_{\rm eq}$ for various
temperatures, where $\t_{\rm eq}$ is the relaxation time determined as the $1/e$
decay time of $\Pi_{\rm eq}(\t)$. 
It is evident that the decay becomes broader with decreasing temperature.
This means that in this model time-temperature superposition does not hold in
equilibrium. 
Without showing the results here, I only mention that the values for the
relaxation time $\t_{\rm eq}$ and the stretching parameter $\b_{\rm eq}$ in a
Kohlrausch fit to $\Pi_{\rm eq}$, 
$\Pi_{\rm eq}(\t)\sim\exp{\left[-(\t/\t_{\rm eq})^{\b_{\rm eq}}\right]}$, 
roughly follow the expressions given by Monthus and Bouchaud\cite{MB96},
$\ln{(\t_{\rm eq})}\sim (1/T)^2$ and $\b_{\rm eq}\sim[1+(a/T)]^{-1/2}$ where $a$
is a constant.

In order to see in which time range aging effects can be observed in
$\Pi(t,t_w)$ for finite $t_w$, in Fig.2a I show $\Pi(t_w+\t,t_w)$ as a function of
the measuring time $\t$ for a quench from infinite temperature, $\b_0=0$, to
various working temperatures.  
It is seen that the decay times $\t(t_w)$ span a much broader range at low
temperatures than at high temperatures. 
A closer look additionally reveals the fact that the decay is broader for small
and large waiting times than for intermediate $t_w$. 
The origin of this behavior lies in the waiting time dependent width of the
distribution of the population, $p(\e,t_w)$, cf. the more detailed discussion on
this point in ref.\cite{G56}. 

The time window in which significant aging effects are to be expected can most
easily be quantified by considering the mean relaxation time,
\[
\lg\t(t_w)\rg=\int\!\!d\t\Pi(t_w+\t,t_w)
\]
which does not depend on the width of the actual decay. 
The mean relaxation times for a quench from $T_0=\infty$ to various temperatures  
are shown in Fig.2b. 
The difference of the mean relaxation time directly after the quench,
$\lg\t_0\rg=\lg\t(0)\rg=\k_\infty^{-1}\exp{(\b^2\s^2/2)}$ and the equilibrium
mean relaxation time 
$\lg\t_{\rm eq}\rg=\lg\t(\infty)\rg=\k_\infty^{-1}\exp{(3\b^2\s^2/2)}$, shown in
the inset in Fig.2b, gives a measure for the 'aging time window' at a given
temperature. 

As is exemplified in Fig.3a, the overall time window for visible aging effects
diminishes not only with increasing final working temperature, but also with
lower initial temperature $T_0$. 
It is evident that the effect of lowering $T_0$ is similar to measuring the
correlation at a higher working temperature for a given $T_0$. 
In Fig.3b the relaxation time $\t(t_w)$ and the stretching parameter $\b(t_w)$ as
obtained from Kohlrausch fits to $\Pi(t_w+\t,t_w)$, 
$\Pi(t_w+\t,t_w)\sim\exp{\left[-(\t/\t(t_w))^{\b(t_w)} \right]}$, are shown. 
One can see that the overall spread in relaxation time diminishes with decreasing
$T_0$ and that $\t(t_w)$ approaches $\t_{\rm eq}$ at the working temperature $T$
on the time scale of $\t_{\rm eq}$ itself. 
This is because there is no other time scale present in the model.
The stretching parameter $\b(t_w)$ first increases and then decreases again as a
function of the waiting time, similar to what has been found earlier in a
free-energy model for the primary relaxation of viscous liquids\cite{G54}. 
Note that in the limit of short and long $t_w$ the $\b(t_w)$-values, 
$\b(0)=\b(\infty)$, are all the same because in the present model the width $\s$
of $\eta(\e)$ and $p^{\rm eq}(\e)$ are independent of temperature.

Next, we turn to a discussion of the FDR, eq.(\ref{X.def}), which is calculated
from the fluctuation-dissipation relation, eq.(\ref{FDT.gen}). 
For long waiting times one finds 
\be\label{X.eq}
X(t,t_w)=\g+\mu\quad\mbox{for}\quad t_w\gg\t_{\rm eq}
\ee
as expected because the system is in equilibrium. 
Furthermore, also the equal-time value yields the same result, 
\be\label{X.t.t}
X(t,t)=\lim_{(t-t_w)\to0}X(t,t_w)=\g+\mu
\ee
This can easily be seen from the fact that for small time-differences one has
$\partial_{t_w}\Pi(t,t_w)=-\partial_t\Pi(t,t_w)$.
For finite $t_w$ and long times, however, the FDR reaches a non-trivial limiting
value
\be\label{X.inf}
X_\infty(t_w)=\lim_{(t-t_w)\to\infty}X(t,t_w)=\g
\ee
independent of $T_0>T$. 
This result is very similar to the one obtained in ref.\cite{G60}. 
It is interesting to compare the result with the corresponding expression for the
trap model with an exponential density of states, where it is found that the long
time limit of $X_\infty(t_w)$ tends towards $\g$,
$X_\infty=\lim_{t_w\to\infty}X_\infty(t_w)$\cite{Ritort03}. 
Thus, apparently, the model with a gaussian density of states resembles some
features of the model with an exponential one for intermediate waiting times.
If interpreted in terms of a time scale dependent effective temperature,
$T/X_\infty(t_w)=1/\g$, one finds that it can be higher ($\g<1$) or smaller
($\g>1$) than the bath temperature $T$.  
It should be mentioned, however, that one typically will have $\g\leq1$. 
For instance, in a 'force model', one would assume the bias induced by the
external field to be symmetric and therefore $\g=\mu=1/2$.

The detailed behavior of the FDR is shown in Fig.4, where I plotted
$X(t_w+\t,t_w)$ versus the measuring time $\t$ for various $t_w$ and $\g=0$ for a
quench from $T_0=\infty$ to $T=0.3\s$. 
It is seen that $X(t_w+\t,t_w)$ starts from the short-time value $X(t_w,t_w)=1$
and decays to its long-time limit, eq.(\ref{X.inf}).
This decay takes place on the time scale $(\t/t_w)\sim1$ as is to be expected for
a model with a single time scale. 
For very long $t_w$, the decay of $X$ will not be observed due to the long
measuring times required. 
If one starts with lower $T_0$ the main difference is that the decay from $X=1$
to $X=\g$ for a given $t_w$ takes place at somewhat later measuring times $\t$.
This is because for smaller $T_0$ the system is closer to equilibrium for a
given $t_w$. 
Thus, one roughly can compare $X$ for the lower $T_0$ and a given $t_w$ with the
corresponding one for the higher $T_0$ but a longer $t_w$.

Instead of computing $X(t,t_w)$ directly, the FDR or the effective temperature is
usually obtained from a so-called fluctuation-dissipation (FD) plot, i.e. a plot
of the integrated response $\chi(t,t_w)=\int_{t_w}^t\!\!dsR(t,s)$ versus the
correlation function $\Pi(t,t_w)$\cite{CKP97,CR03}. 
The FDR then is extracted as the slope in such a FD-plot. 
It has been pointed out earlier that such plots have to be constructed with some
care in the general case\cite{FS02}. 
In some situations a plot of $\chi(t_w+\t,t_w)$ vs. $\Pi(t_w+\t,t_w)$ with $\t$ as
the curve parameter and fixed $t_w$ does not necessarily give the correct FDR. 
This is only ensured if $t_w$ is used as the curve parameter due to the fact that
according to the definition of the integrated response one has 
$R(t,t_w)=-{\partial\chi(t,t_w)\over\partial_{t_w}}$. 
The reason is that in general the different derivatives $\partial_{t_w}\Pi(t,t_w)$
and $\partial_t\Pi(t,t_w)$ cannot be interchanged. 
For some models with transient aging behavior, it has indeed been found that the
former construction yields wrong results\cite{G59,G60}. 
Also for the gaussian trap model, FD-plots with $\t$ as the curve parameter yield
wrong results for the FDR. 
This fact is exemplified in Fig.5a, where such FD-plots for different temperatures
and $t_w=10^{-15}\t_{\rm eq}$ are shown.  
The initial temperature was chosen as $T_0=\infty$. 
The inset shows the temperature dependence of the slopes extracted from these
plots via a linear regression in the interval $0.3<\Pi<0.7$.
Note that all slopes are larger than $X_\infty(t_w)=\g=0.5$. 
The situation is very similar to the one in a free-energy model for
glassy relaxation discussed in detail in ref.\cite{G60}. 

In Fig.5b, a FD-plot with $t_w$ as the curve parameter and various values of the
time $t$ is shown for $\g=0$ (upper panel) and $\g=1/2$ (lower panel).
It is evident that in this case the limiting slopes coincide with
$X_\infty(t_w)$, eq.(\ref{X.inf}). 
However, it is also clear that the curves change continuously until this limiting
slope is reached. 
In a strict sense this would mean that a thermometer would measure an increasing
effective temperature as a function of time even though in the model there is only
a single time scale. 
Therefore, similar to the case of the trap model with an exponential density of
states, it appears that only the limiting slope $X_\infty(t_w)$ may be a useful
candidate for the definition of an effective temperature\cite{FS02}. 
Additionally, the detailed behavior of the curves also depends on the initial
temperature $T_0$, as can be seen from Fig.5c which shows FD-plots for various
$T_0$ and a fixed time $t=10^{-4}\t_{\rm eq}$ for $\g=0$, $\mu=1$.
The limiting value for the slope is the same, as already pointed out above.

From the above discussion it appears that the quantity $T/X_\infty(t_w)$ may serve
as a $t_w$-dependent effective temperature in the gaussian trap model. 
As already noted in the previous section, there are other possible definitions of
effective temperatures, in particular the one given in eq.(\ref{Teff.om}),
$T_{\rm eff}(\om,t_w)={\om C(\om,t_w)\over \chi''(\om,t_w)}$. 
Here, the natural way to define the Fouriertransform is with respect to the
measuring time\cite{GR05,RS03}: 
\Be\label{C.chi.om.t_w}
C(\om,t_w)=&&\hspace{-0.6cm}{\rm Re}\int_0^\infty\!d\t\Pi(t_w+\t,t_w)e^{i\om\t}
\nonumber\\
\chi''(\om,t_w)=&&\hspace{-0.6cm}
                    {\rm Im}\int_0^\infty\!d\t R(t_w+\t,t_w)e^{i\om\t}
\Ee
which is similar to the definition used in most experiments.
It can be shown analytically that $T_{\rm eff}(\om,t_w)$ defined this way has
the following limiting behavior:
\be\label{Teff.eq.limits}
\lim_{t_w\to\infty}T_{\rm eff}(\om,t_w)=
\lim_{\om\to\infty}\lim_{t_w\to 0}T_{\rm eff}(\om,t_w)={T\over\mu+\g}
\ee
as expected. 
However, the low-frequency limit is given by:
\be\label{Teff.om.0}
\lim_{\om\to 0}\lim_{t_w\to 0}T_{\rm eff}(\om,t_w)={T\over\mu+\g e^{2(\s/T)^2}}
\ee
which is different from $T/X_\infty(t_w)$. 
For $\mu=1$ and $\g=0$ no deviations from the equilibrium FDT can be observed at
all. 
When considered as a function of frequency, $T_{\rm eff}(\om,t_w)$ starts from
the low-frequency limit $T/(\mu+\g e^{2(\s/T)^2})$ and roughly at 
$\om t_w\sim1$ smoothly crosses over to the high-frequency limit $T/(\mu+\g)$. 
The main difference to $T/X$ is that the low-frequency limit usually is much
smaller than the bath temperature $T$.
An obvious reason for this discrepancy lies in the definition of the
Fouriertransform. 
The definition of the FDR involves only the partial derivative
$\partial_{t_w}\Pi(t,t_w)$, whereas $\om C(\om,t_w)$ according to
eq.(\ref{C.chi.om.t_w}) corresponds to the the Fouriertransform of
$\partial_t\Pi(t,t_w)$. 
Thus, the situation appears to be very similar to the one met when considering the
different ways of constructing FD-plots discussed above. 
There is, however, one difference. 
For the present model it is not at all clear that a definition of an effective
temperature via the Fouriertransform of $\Pi(t,t_w)$ and $R(t,t_w)$ is related to
the FDR in a direct manner. 
This is due to the fact that one has to consider the Fouriertransform of 
$R(t,t_w)=\b X(t,t_w)\partial_{t_w}\Pi(t,t_w)$ which in general will give a
convolution instead of a simple product.

It is tempting to use an alternative definition of the
Fouriertransform\cite{CKP97}, 
$\hat C(\om,t)={\rm Re}\int_0^t\!ds\Pi(t,s)e^{-i\om(t-s)}$ and
$\hat \chi''(\om,t)={\rm Im}\int_0^t\!ds R(t,s)e^{-i\om(t-s)}$,
for which one can see that $\om\hat C(\om,t)$ is related to the Fouriertransform
of $\partial_{t_w}\Pi(t,t_w)$. 
One then has $\hat T_{\rm eff}(\om,t)={\om\hat C(\om,t)\over\hat\chi''(\om,t)}$. 
However, this definition implies that one considers the situation $\om t\gg 1$,
in order for the oscillatory part of the functions involved to be negligible. 
As one can show analytically that $\hat T_{\rm eff}(\om,t\to\infty)
=\hat T_{\rm eff}(\om\to\infty,t)=T/(\g+\mu)$ this condition means that the
relevant small frequency-limit is hard to reach in practice, in particular
because the relaxation takes place on the scale $\om t\sim1$.
\section*{IV. Conclusions}
In the present paper I considered the aging behavior of the correlation function
and the linear response in a gaussian trap model.
As in this model equilibrium is reached for long waiting times after a quench
($t_w\to\infty$), all aging effects are of a transient nature. 
The time window in which aging effects can be observed expands with decreasing
working temperature $T$ if a quench from a fixed high temperature $T_0$ is
considered.
If $T$ is kept fixed and $T_0$ is increased the 'aging time window' also
increases. 

If the waiting time is chosen to be longer than the equilibrium relaxation time
($1/e$ decay time), $t_w\gg\t_{\rm eq}$, the two-time quantities like the
correlation function $\Pi(t,t_w)$ and the response function $R(t,t_w)$ approach
their equilibrium values and only depend on the time-difference $(t-t_w)$. 
In this situation the FDT is obeyed. 
For short waiting times, on the other hand, one has an explicit dependence of
$\Pi(t,t_w)$ and $R(t,t_w)$ on both times and the functions no longer are
time-translational invariant. 
In this situation strong violations of the FDT are observed. 
In particular, for $t_w<\t_{\rm eq}$ the FDR $X(t,t_w)$ tends to a non-trivial
long-time limit $X_\infty(t_w)=\g$, cf. eq.(\ref{X.inf}), independent of $T_0>T$. 
The constant $\g$ is determined by the coupling of the system's dynamics to an
external field via the transition rates $W^{(H)}(\e'|\e)$, cf. eq.(\ref{kap.h}). 
The only possible temperature dependence of $X_\infty(t_w)$ stems from a
T-dependence of $\g$. 
The value for $X_\infty(t_w)$ coincides with that found for the trap model with
an exponential density of states\cite{Ritort03}, albeit for long waiting times. 

The FDR can be used for the definition of a time scale dependent effective
temperature, $T/X(t,t_w)$. 
Usually, the effective temperature is extracted from FD-plots, which can be
constructed in different ways.
As has been pointed out\cite{FS02}, the correct way to construct FD-plots is to
use $t_w$ as the curve parameter for fixed $t$. 
If FD-plots are constructed this way, the limiting slope coincides with
$X_\infty(t_w)$. 
The fact that these plots are curved indicates that not the function $X(t,t_w)$ 
as a whole but only the limiting value $X_\infty(t_w)$ may be used for a
meaningful definition of an effective temperature. 
If alternatively FD-plots with the measuring time as curve parameter for
different $t_w<\t_{\rm eq}$ are considered one finds that one can define a slope
in these plots which, however, does not coincide with $X(t,t_w)$ and furthermore
show some artificial temperature dependence.
A similar problem occurs when the definition of an effective temperature in the
frequency domain is considered. 
The usual definition of the Fouriertransform, eq.(\ref{C.chi.om.t_w}), yields an
effective temperature that is distinct from what one finds from the FDR.
In this case, similar as for the FD-plots, one reason is that one cannot
interchange the derivatives with respect to the earlier and the later time in the
correlation function.

In the recent past FD-plots have been constructed from computer simulations of
model supercooled liquids\cite{barrat03, KB99, SVST03}. 
For the model of a fragile liquid one finds that the FDT is obeyed for large
values of the correlation and then the slope changes to a smaller value indicating
an effective temperature larger than the bath temperature\cite{barrat03, KB99}. 
For a strong liquid, however, one finds an effective temperature smaller than the
bath temperature\cite{SVST03}. 
These plots have been constructed with the measuring time as the curve parameter
and fixed waiting times.

Given the fact that the gaussian trap model apparently is able to capture some
important features of the dynamics observed in computer simulations on
supercooled liquids, it is tempting to speculate that FD-plots with the waiting
time as the curve parameter might look different to the ones published so far. 
Therefore, it would be interesting to calculate FD-plots from simulation data the
correct way and compare the results to the previous findings. 
\section*{Acknowledgment}
I thank Roland B\"ohmer and Jeppe Dyre for helpful discussions on the subject.
The paper is dedicated to Hans Sillescu on occasion of his 70'th birthday.

\newpage
\section*{Figure captions}
\begin{description}
\item[Fig.1 : ]
$\Pi_{\rm eq}(\t)$ versus $\t$ for various temperatures. 
\item[Fig.2 : ]
{\bf a:} $\Pi(t_w+\t,t_w)$ versus measuring time $\t$, scaled to the
equilibrium relaxation time $\t_{\rm eq}$ (1/e decay time) for a quench from
$T_0=\infty$.\\ 
{\bf b:} $\lg\t(t_w)\rg$ versus $t_w/\lg\t_{\rm eq}\rg$ showing the time window in
which aging effects are to be expected. 
The inset shows the mean relaxation times 
$\lg\t_0\rg=\lg\t(0)\rg$ and $\lg\t_{\rm eq}\rg=\lg\t(\infty)\rg$ as a function
of inverse temperature.
\item[Fig.3 : ]
{\bf a:} $\Pi(t_w+\t,t_w)$ versus $\t /\t_{\rm eq}$ for quenches from various
$T_0=0$, $\s$, $0.5\s$ and $\log_{10}(t_w/\t_{\rm eq})=-15,-5,-3,-1,1,5$.\\ 
{\bf b:} $\t(t_w)$ (upper panel) and $\b(t_w)$ (lower panel) versus $t_w/\t_{\rm
eq}$ as obtained from fits to a Kohlrausch function,
$\exp{\left[-(\t / \t(t_w))^{\b(t_w)} \right]}$.
\item[Fig.4 : ]
The FDR $X(t_w+\t,t_w)$ versus $\t /\t_{\rm eq}$ for a quench from $T_0=\infty$ to
$T=0.3\s$ and $\g=0$, $\mu=1$ for various waiting times $t_w$.
\item[Fig.5 : ]
{\bf a:} Plot of $T\chi(t_w+\t,t_w)$ versus $\Pi(t_w+\t,t_w)$ (FD-plot) with the
measuring time $\t$ as the curve parameter and fixed waiting time,
$t_w=10^{-15}\t_{\rm eq}$, for a quench from $T_0=\infty$ to 
$T=0.2,0.3,0.4,0.5,0.6,1.0,3.0\s$ from bottom to top. Here, $\g=\mu=1/2$.
The inset shows the slopes extracted from these plots via a linear regression for
the data in the interval $0.3<\Pi<0.7$. 
\\
{\bf b:} Plot of $T\chi(t,t_w)$ versus $\Pi(t,t_w)$ (FD-plot) for a quench from
$T_0=\infty$ to $T=0.25\s$ and times $\log_{10}(t/\t_{\rm eq})=-6,-4,-2,0,6$ from
bottom to top. 
In the upper panel $\g=0$, $\mu=1$ has been chosen, i.e. $X_\infty(t_w)=0$ and in
the lower panel $\g=\mu=1/2$ ($X_\infty(t_w)=1/2$, dotted line).\\
{\bf c:} $T\chi(t,t_w)$ versus $\Pi(t,t_w)$ for a quench from $T_0=\infty$,
$1.0\s$, $0.5\s$, and $0.4\s$ to $T=0.25\s$ and $t/\t_{\rm eq}=10^{-4}$, $\g=0$,
$\mu=1$. 
%
%
\end{description}
\newpage
\begin{figure}
\includegraphics[width=15cm]{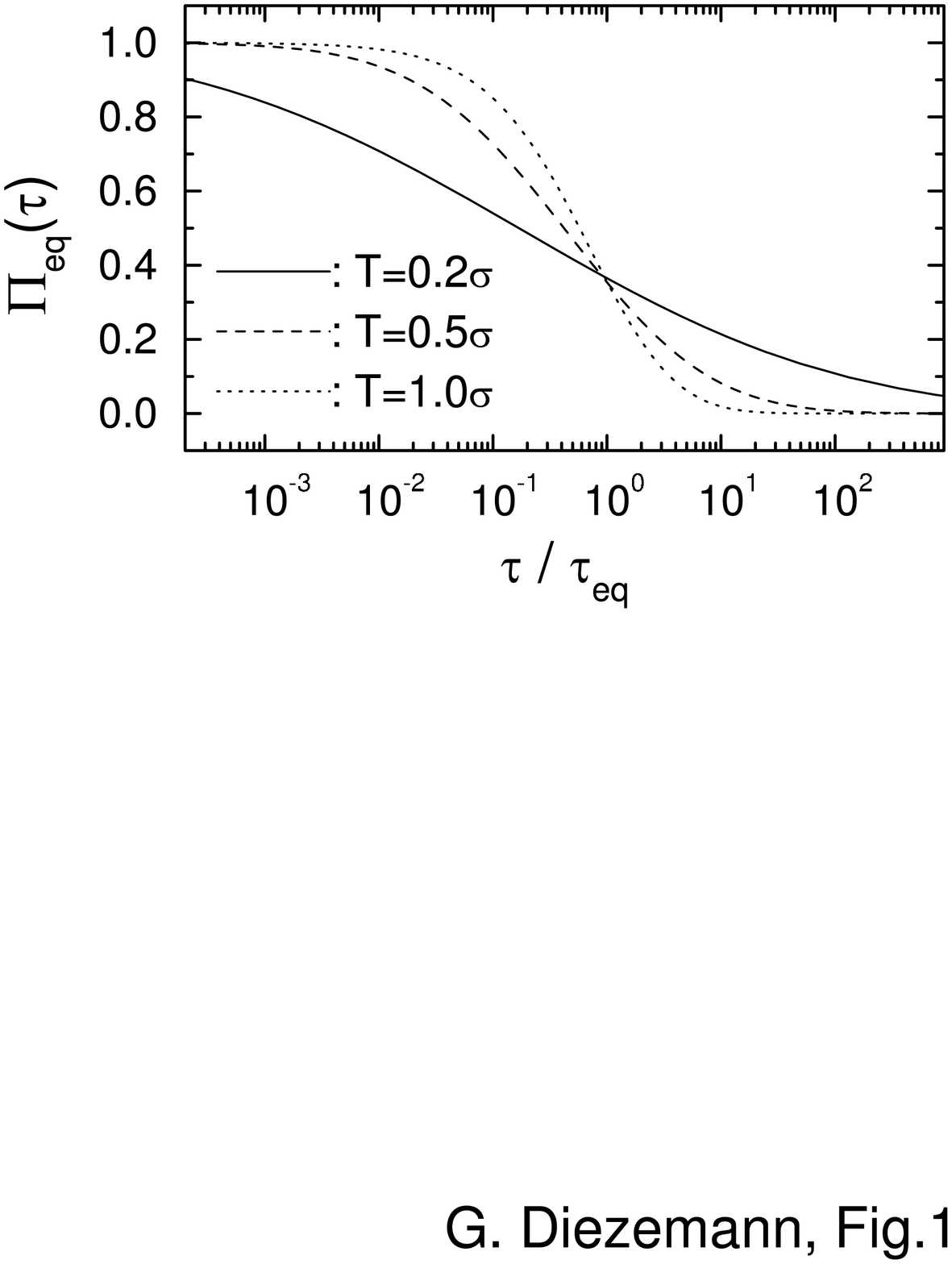}
\end{figure}
\newpage
\begin{figure}
\includegraphics[width=15cm]{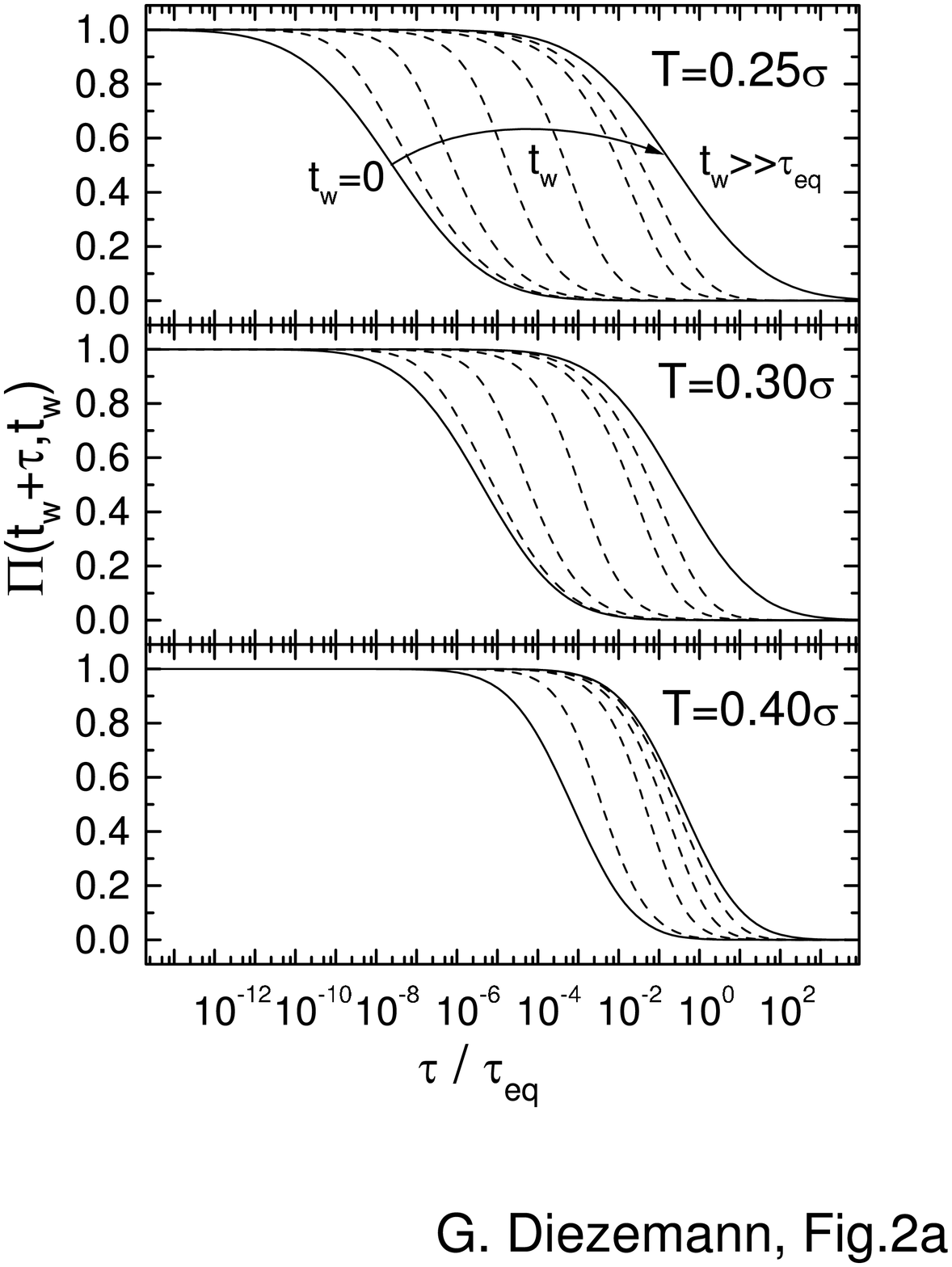}
\end{figure}
\newpage
\begin{figure}
\includegraphics[width=15cm]{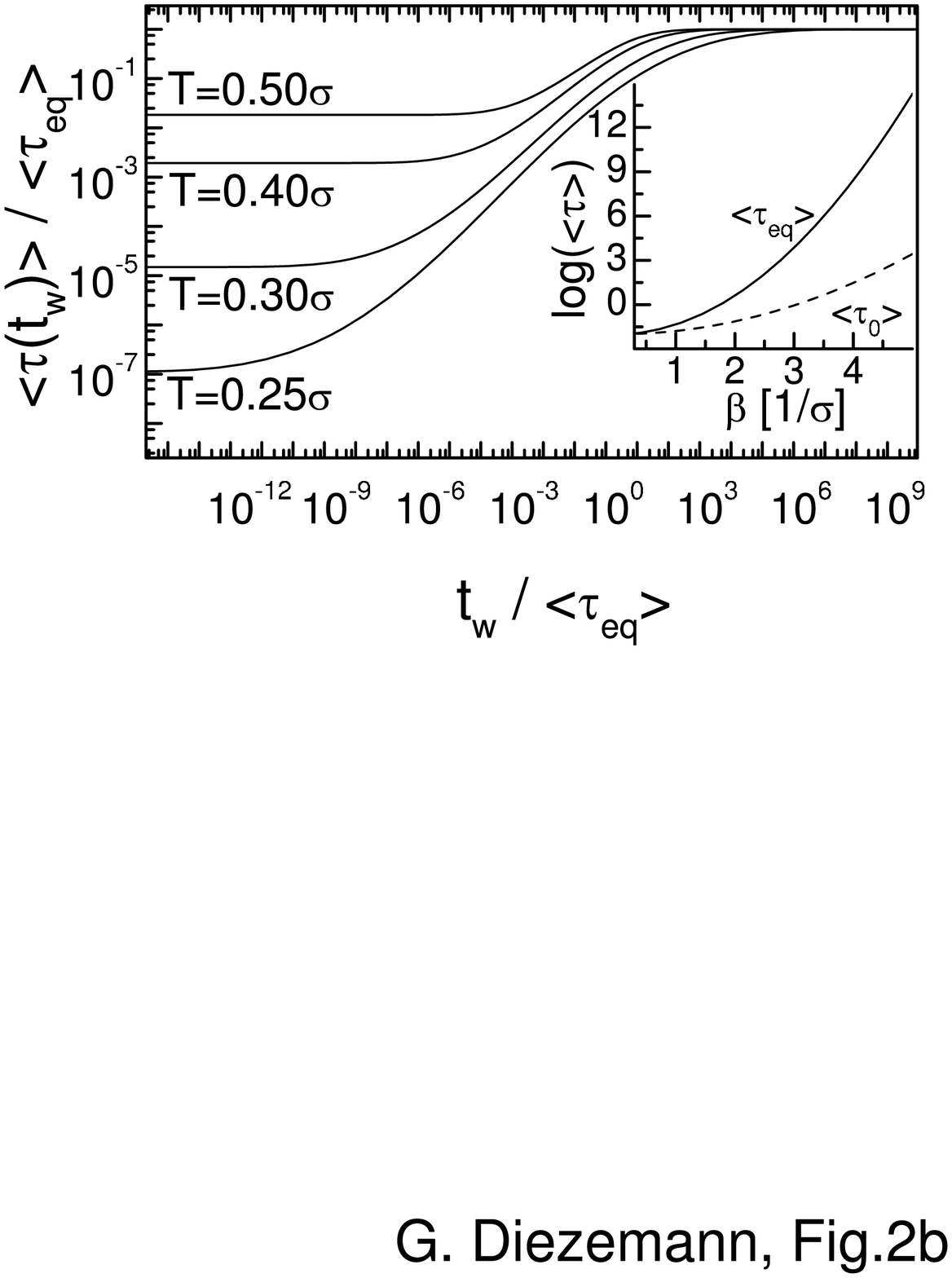}
\end{figure}
\newpage
\begin{figure}
\includegraphics[width=15cm]{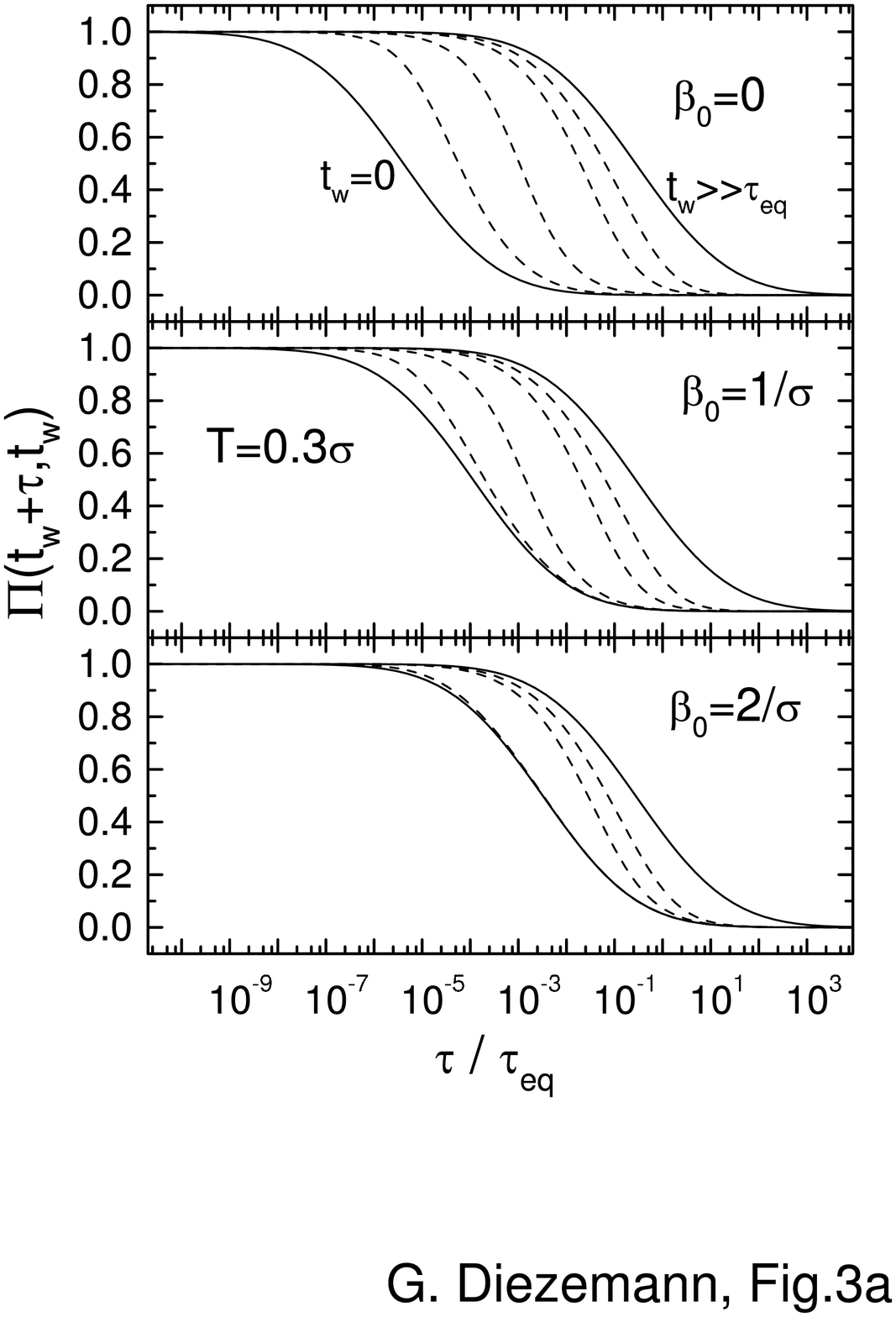}
\end{figure}
\newpage
\begin{figure}
\includegraphics[width=15cm]{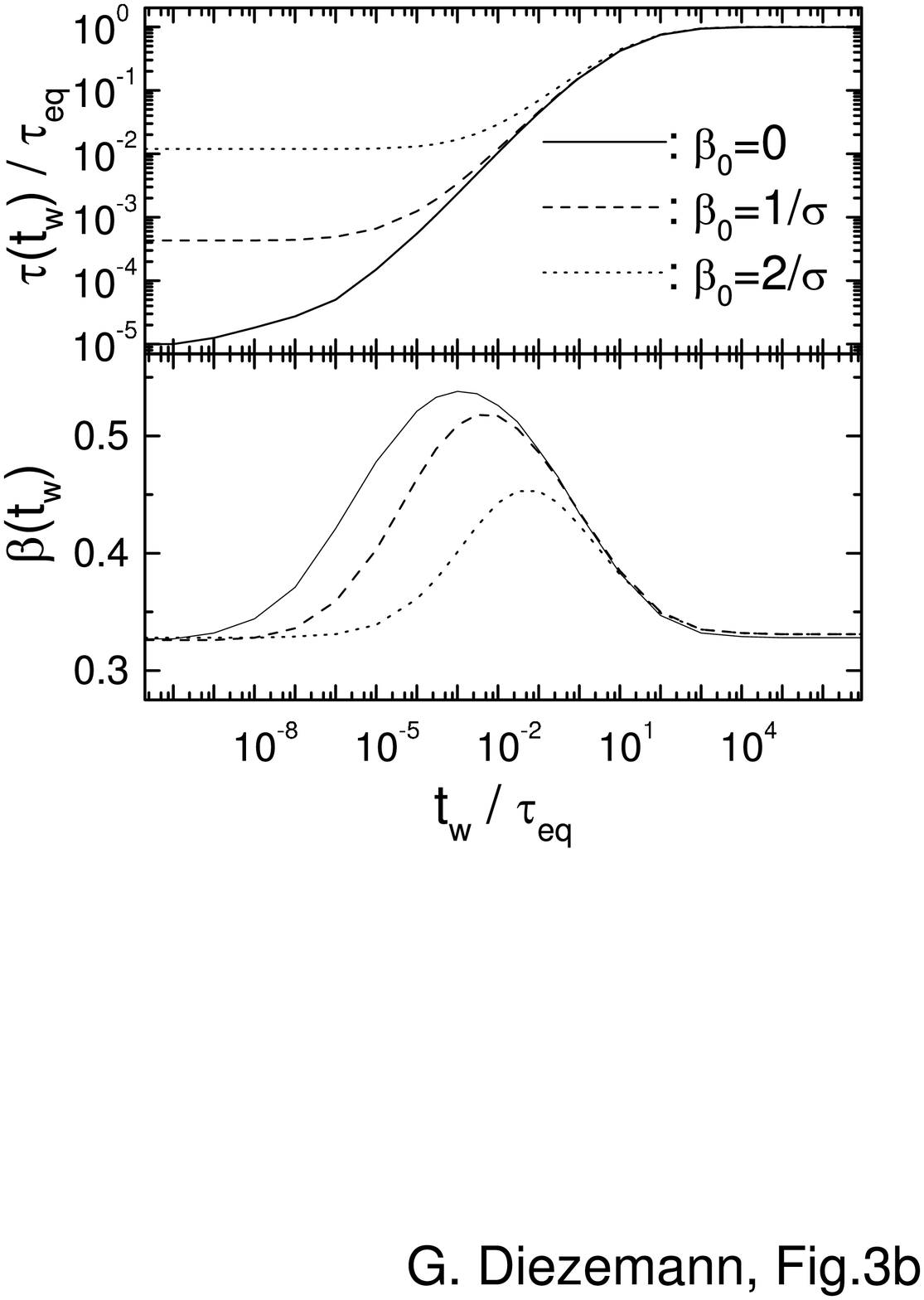}
\end{figure}
\newpage
\begin{figure}
\includegraphics[width=15cm]{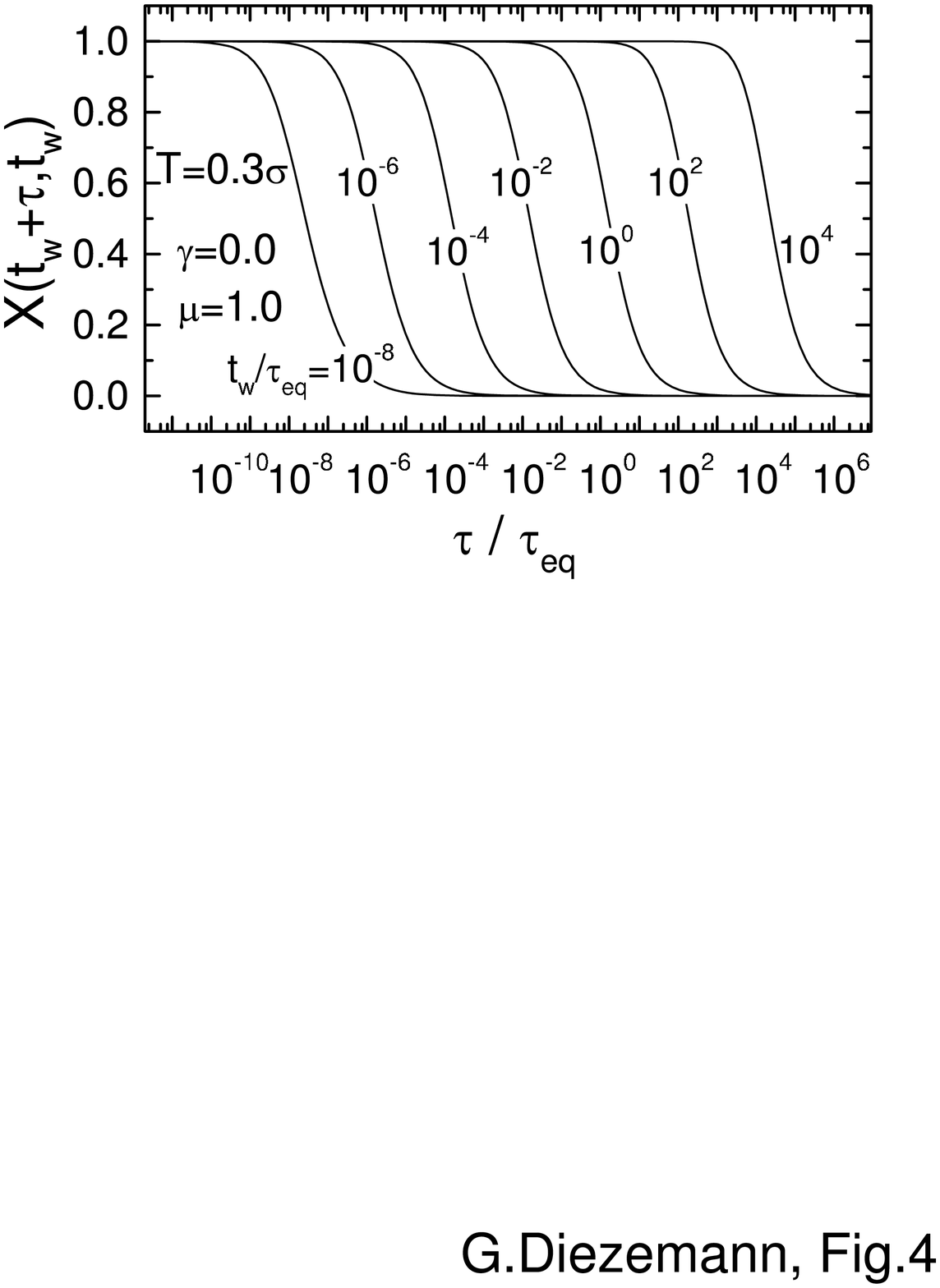}
\end{figure}
\newpage
\begin{figure}
\includegraphics[width=15cm]{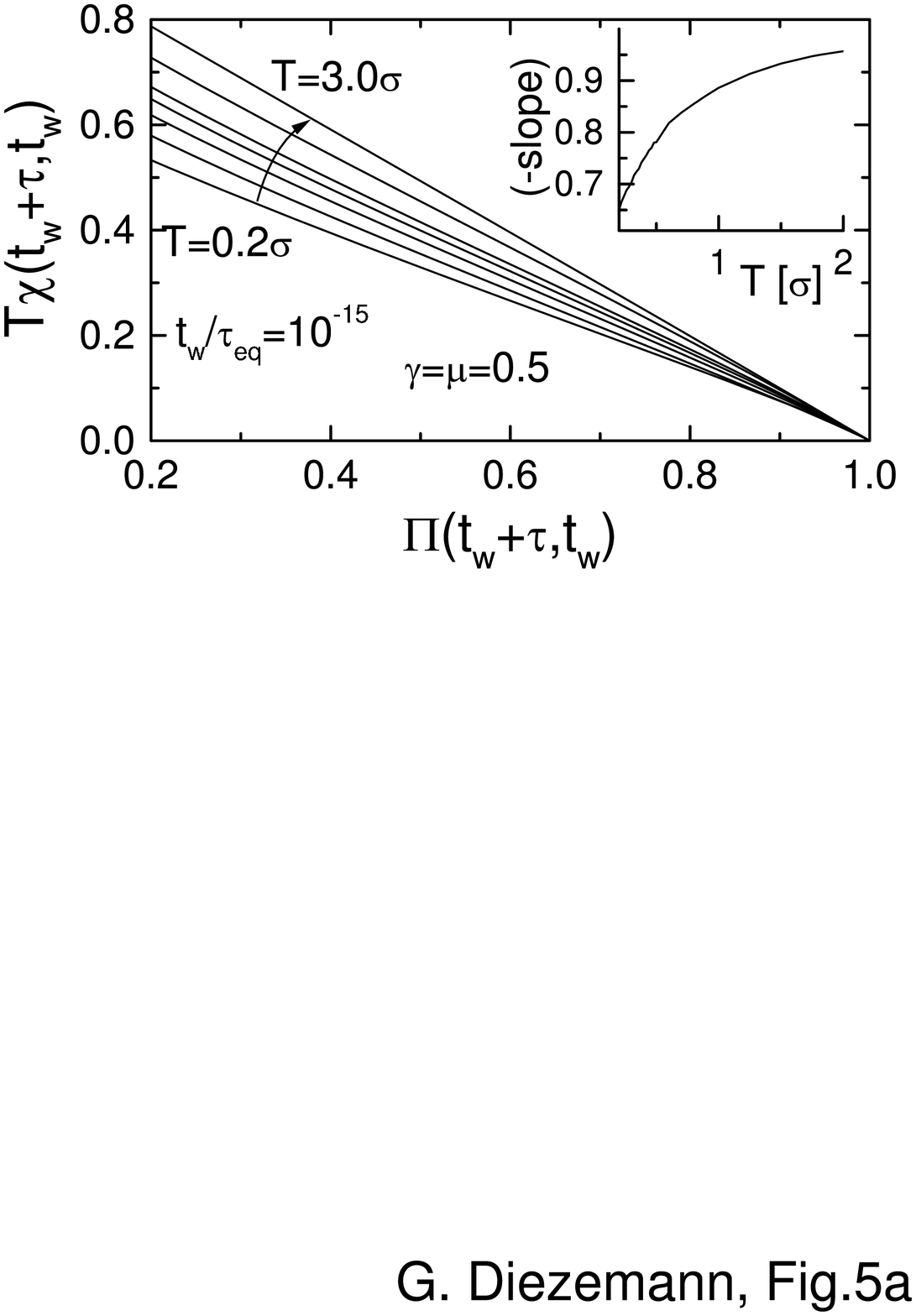}
\end{figure}
\newpage
\begin{figure}
\includegraphics[width=15cm]{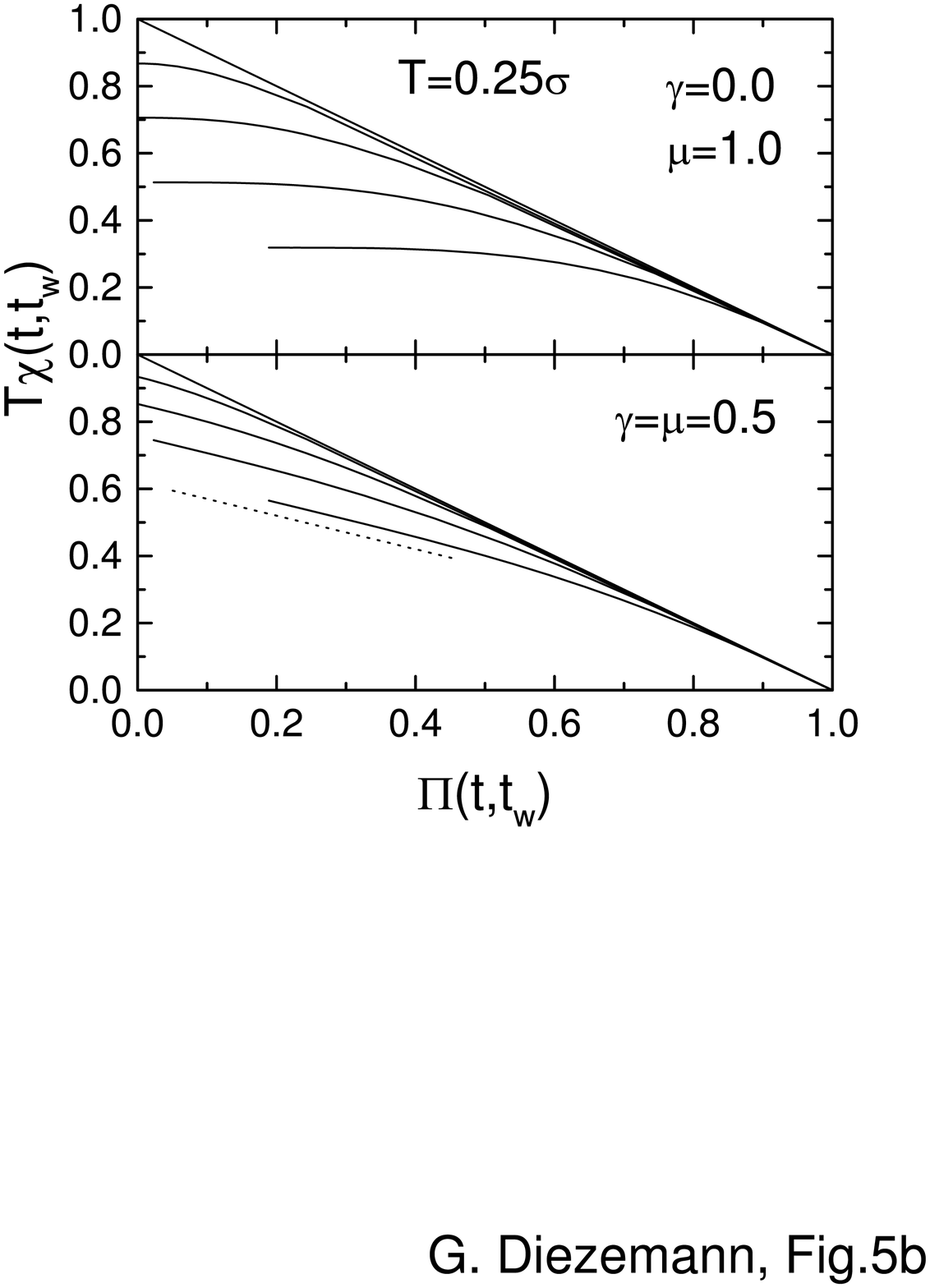}
\end{figure}
\newpage
\begin{figure}
\includegraphics[width=15cm]{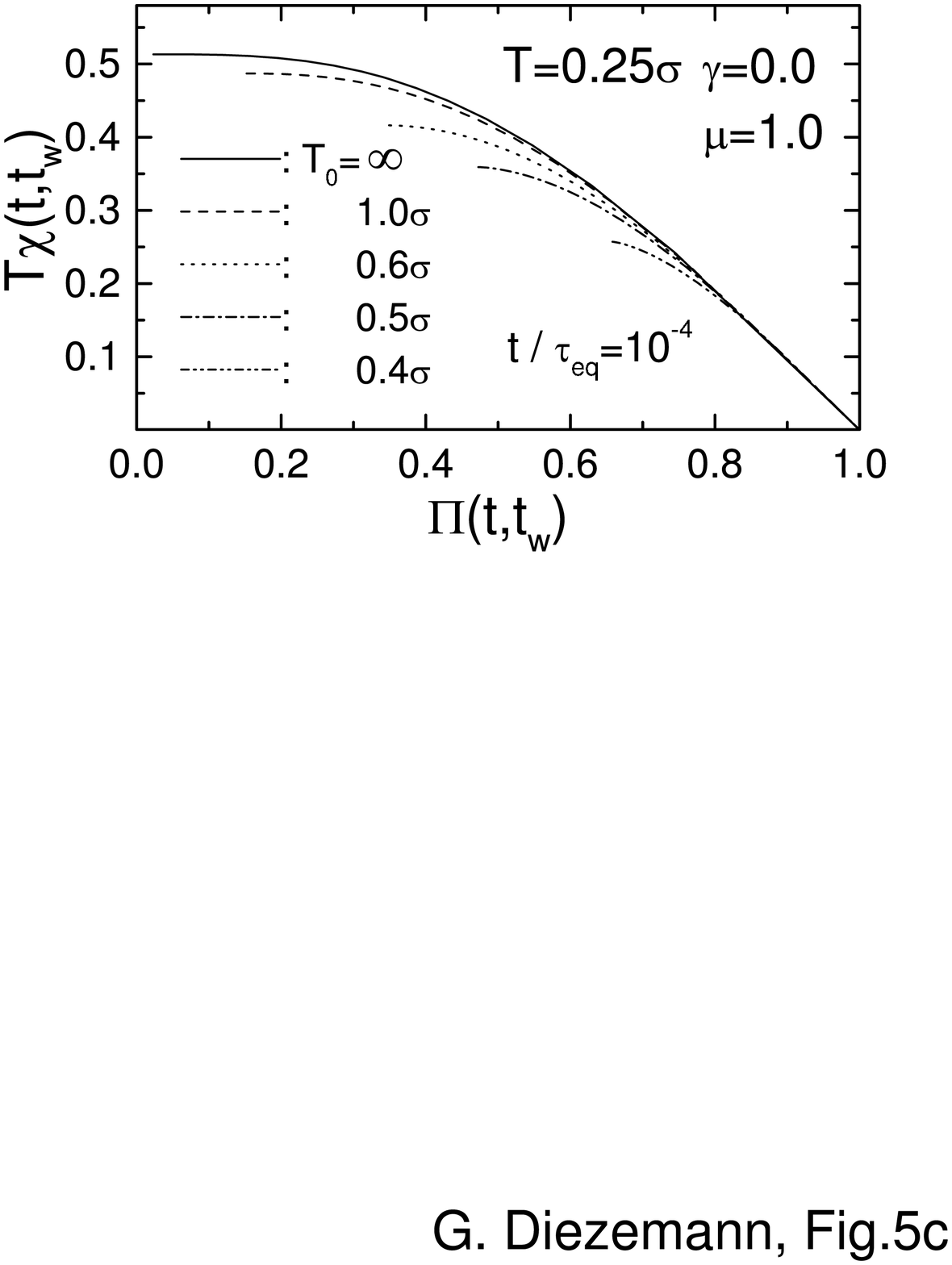}
\end{figure}
\end{document}